\documentclass[preprint,notoc]{JHEP3}
\usepackage{epsfig}

\def\eslt{\not\!\!{E_T}}
\def\to{\rightarrow}

\def\bi{\begin{itemize}}
\def\ei{\end{itemize}}
\def\hhat{\hat{h}}
\def\CL{{\cal {L}}}
\def\th{\theta}

\def\tf{\tilde f}

\def\tst{\tilde t}

\def\tg{\tilde g}

\def\tell{\tilde\ell}
\def\tq{\tilde q}
\def\tw{\widetilde W}
\def\tz{\widetilde Z}
\def\alt{\stackrel{<}{\sim}}
\def\agt{\stackrel{>}{\sim}}

\title{Radiative Neutralino Decay in Supersymmetric Models}

\author{Howard Baer and Tadas Krupovnickas
\\ Department of Physics, Florida State University\\ 
Tallahassee, FL, USA 32306\\
E-mail: \email{baer@hep.fsu.edu}, \email{tadas@hep.fsu.edu}}

\preprint{\vbox{\hbox{FSU-HEP-020530}}}

\abstract{The radiative decay $\tz_2\to \tz_1\gamma$ 
proceeds at the one-loop level in the MSSM. It can be the
dominant decay mode for the second lightest neutralino $\tz_2$
in certain regions of parameter space of supersymmetric models, where 
either a dynamical and/or kinematic enhancement of the branching fraction 
occurs. 
We perform an updated numerical study of this decay mode in both the 
minimal supergravity model
(mSUGRA) and in the more general MSSM framework.
In mSUGRA, the largest rates are found in the
``focus point'' region,
where the $\mu$ parameter becomes small, and the lightest neutralinos
become higgsino-like; in this case, radiative branching fraction can reach the
1\% level. 
Our MSSM analysis includes a scan over independent positive and 
negative gaugino masses. We show branching fractions can reach the 10-100\%
level even for large values of the parameter $\tan\beta$. These regions
of parameter space are realized in supergravity models with non-universal
gaugino masses. 
Measurement of the radiative neutralino branching fraction may help
pin down underlying parameters of the fundamental supersymmetric model.
}


\begin{document}

\section{Introduction}
\label{sec:intro}

One of the major goals of collider experiments is to verify or disprove
the existence of weak scale supersymmetric matter\cite{reviews}. 
Given sufficient energy and integrated luminosity, sparticles
should be produced at large rates, if indeed they exist. Once produced,
sparticles will decay via a cascade into other sparticles until the
state containing the lightest SUSY particle (LSP) is reached\cite{cascade}.
If kinematically accessible, tree level two-body decays will
comprise the dominant sparticle branching fractions. If two-body
modes are suppressed or closed, then tree level three-body decays 
can dominate. It is also possible for sparticles to decay
into two-body final states where the decay is mediated by loop
diagrams. Important examples include $\tg\to g\tz_i$, $\tz_i\to\tz_j\gamma$
($i,j=1-4$ with $i>j$) 
and $\tst_1\to c\tz_1$ for sparticles, and $h\to \gamma\gamma$
for the lightest SUSY Higgs scalar.

The radiative decay $\tz_i\to\tz_j\gamma$ has been examined previously 
in a number of papers, since it can give rise to a distinctive
energetic, isolated photon plus missing energy signature.
The decay proceeds through loops containing intermediate
charged scalars and charginos. After several approximate
calculations\cite{early}, the complete calculation was presented
in both the non-linear and linear $R_\xi$ gauge
by Haber and Wyler\cite{haber}. These authors note that the radiative
neutralino decay branching fraction may be large 
in MSSM parameter space regions
where $M_1,\ M_2,\ |\mu |\ll M_Z$ (light neutralino radiative decay), or when 
$|\mu |,M_Z\ll M_1,\ M_2\sim $ TeV (higgsino to higgsino radiative decay). 

Detailed numerical studies of
the $\tz_2\to\tz_1\gamma$ decay rate have been performed by
Ambrosanio and Mele\cite{mele}. These authors found that the
$\tz_2\to\tz_1\gamma$ decay rate can in fact be the dominant
$\tz_2$ decay mode in regions of MSSM model parameter space
where tree level two-body $\tz_2$ decay modes are not allowed, and 
the competing $\tz_2$ three-body decay modes are {\it dynamically} 
and/or {\it kinematically} suppressed. 

At low $\tan\beta$, the three-body neutralino decay $\tz_2\to\tz_1 f\bar{f}$
occurs dominantly via five Feynman diagrams involving virtual sfermion
and $Z$ boson exchange. 
The dynamical suppression of the three-body
decay modes occurs when one of $\tz_2$ and $\tz_1$ is gaugino-like, while the 
other is higgsino-like.
Since the $\tz_i f\tf$ vertex contains the gaugino component of the
neutralinos, the decay via a virtual sfermion $\tf^*$ is 
suppressed if one (or both) 
of $\tz_2$ or $\tz_1$ is higgsino-like.
In addition, 
the $Z\tz_2\tz_1$ coupling contains the higgsino
components of the neutralinos so that if one of $\tz_1$ and/or $\tz_2$ is
gaugino-like, then decay via the $Z^*$ diagram is suppressed.
The region of MSSM parameter space where dynamical 
suppression of three-body decay modes occurs is when
the gaugino masses $M_1\sim M_2$, $\tan\beta \sim 1$ and $\mu <0$
with $|\mu |$ being sufficiently small. 
In light of recent limits from LEP2 that $m_{\tw_1}>103.5$ GeV and 
$m_h>114.1$ GeV (for a SM-like $h$ boson), the region of
dynamical suppression is largely excluded, since $m_h$ is quite light at 
very low values of $\tan\beta$.

In addition, there can exist
a region of {\it reduced} dynamical suppression. In this case, 
three-body decays via $Z^*$ can be suppressed if one (or both) 
of the neutralinos
is gaugino-like (thus suppressing $\tz_2$ decay via $Z^*$), while
the sfermions are quite heavy (thus suppressing decay through $\tf^*$).
This interesting case occurs naturally as we shall see in the
``focus-point'' scenario\cite{feng} 
of the minimal supergravity (mSUGRA or CMSSM) model\cite{sugra}.

Finally, there can exist regions of kinematical suppression of
three-body $\tz_2$ decays. In this case, $m_{\tz_2}\sim m_{\tz_1}$ so that
$\tz_2\to\tz_1f\bar{f}$ is suppressed kinematically by a factor
$\epsilon^5$, where $\epsilon =(1-m_{\tz_1}/m_{\tz_2})$. The
radiative decay $\tz_2\to\tz_1\gamma$ is suppressed by $\epsilon^3$, so that
as the $m_{\tz_2}-m_{\tz_1}$ mass gap approaches zero, the radiative decay 
increasingly dominates. The kinematic suppression region occurs
when $\tan\beta \sim 1$ {\it or} when $M_1\sim M_2$. In the region
of parameter space where kinematical suppression occurs, however,
the small mass gap between neutralinos also means that the
final state photon energy will be small, unless enhanced by a Lorentz boost
if $\tz_2$ is produced either directly or via cascade decay with 
a high velocity.

The radiative neutralino decays have received some recent attention
by Kane et al.\cite{kane} in attempts to explain an $ee\gamma\gamma$ event
seen by the CDF experiment in run 1 of the Fermilab Tevatron\cite{cdf}.
These ``higgsino-world'' scenarios seem to need $\tan\beta \sim 1$,
$\mu <0$ and a light top squark. These scenarios seem largely
excluded now by LEP2 limits on the masses of the Higgs bosons, which 
require $\tan\beta >2-3$.

In this paper, we re-examine the radiative neutralino decay branching fraction
as a function of supersymmetric model parameter space.
Our calculations include a number of improvements over
previous analyses. 
\begin{enumerate}
\item  We implement the complete $\tz_2\to\tz_1\gamma$
decay calculation into the event generator ISAJET v7.64\cite{isajet}. 
Our calculation
includes full third generation mixing effects amongst the squarks and sleptons.
\item Our treatment of neutralino three-body decays includes the effect
of all third generation Yukawa couplings and sfermion mixings 
in the decay calculations\cite{ltanb}. 
These are especially important at large values of the parameter
$\tan\beta$, where the Yukawa couplings 
$f_b$ and $f_\tau$ can become large. In addition, along with the five
$\tz_2\to\tz_1 f\bar{f}$ decay diagrams proceeding via sfermion and $Z$ boson
exchange, we include
decays through $h$, $H$ and $A$ bosons.
These diagrams can be important, especially at large 
$\tan\beta$\cite{ltanb,bartl,djouadi}.
\item We include the latest constraints from LEP2. As noted before, 
these constraints eliminate many of the regions of parameter space
where $\tz_2\to\tz_1\gamma$ can be large, especially around 
$\tan\beta \sim 1$. 
\item We present results for the paradigm mSUGRA model. We also present
a scan of MSSM parameter space which includes 
negative as well as positive gaugino masses. We find the radiative decay
branching fraction can be large in a particular SUGRA model with
non-univeral gaugino masses. 
\item Finally, in our numerical scans, 
we take note of parameter space regions where the
radiative neutralino branching fraction is {\it not} dominant, but may be
nonetheless observable. For instance, at the CERN LHC, many millions
of sparticle production events may be recorded each year, provided
sparticles are light enough. In this case, branching fractions
of a few per cent or even less may be interesting, and allow one to
measure the neutralino radiative decay branching fraction. Since
the $\tz_2\to\tz_1\gamma$ branching fraction is very sensitive to
parameter space, its measurement may help determine or constrain
some of the fundamental SUSY model parameters.
\end{enumerate}

The rest of this paper is organized as follows. In Sec. \ref{sec:msugra},
we examine the radiative neutralino branching fraction in the mSUGRA
model. We find it can reach levels approaching 1\%, especially in the
well-motivated focus-point region. In Sec. \ref{sec:mssm},
we examine the radiative decay branching fraction in the MSSM. In this case, 
we find regions of parameter space where the branching fraction can approach
50-100\%, even for large values of $\tan\beta$. These regions occur
dominantly where $M_1\sim M_2$ or when $M_1\sim -M_2$. 
We note in Sec. \ref{sec:nusug} that the latter MSSM regions of
enhancement naturally occur in a particular SUGRA model with non-universal
gaugino masses. We map out the interesting regions of parameter space
where the radiative decay branching fraction can be large.
Finally, in an appendix, we present our formulae for radiative neutralino
production, cast in a form consistent with the ISAJET event
generator program.

\section{$\tz_2\to\tz_1\gamma$ branching fraction in mSUGRA model}
\label{sec:msugra}

Our first goal is to examine the radiative neutralino branching fraction 
in the paradigm mSUGRA (or CMSSM) model. Most analyses of supersymmetric
phenomena take place within this model, so it is convenient to know the
rates for radiative neutralino decay in relation to other expected
phenomena. In the mSUGRA model, the MSSM is assumed to be valid between
the grand unified scale ($M_{GUT}\simeq 2\times 10^{16}$ GeV) and the
weak scale ($\sim 1$ TeV). At $M_{GUT}$, it is assumed all scalar
masses are unified to the parameter $m_0$, while all gaugino masses unify to
$m_{1/2}$. In addition, trilinear soft SUSY breaking masses unify to $A_0$.
The gauge and Yukawa couplings and soft SUSY breaking parameters
evolve from $M_{GUT}$ to $M_{weak}$ according to the well-known MSSM
renormalization group equations (RGEs). 
At $M_{weak}$ it is assumed electroweak
symmetry is broken radiatively, which fixes the magnitude of the 
superpotential $\mu$ parameter, and allows one to effectively 
trade the bilinear soft breaking parameter $B$ for the ratio of Higgs vevs
$\tan\beta$. Thus the parameter space is characterized by
\begin{equation} 
m_0,\ m_{1/2},\ A_0,\ \tan\beta\ \ {\rm and}\ \ sign(\mu ),
\end{equation}
where we take the top quark pole mass $m_t=175$ GeV.
In our mSUGRA model calculations, we utilize the iterative RGE
solution embedded within the program ISAJET v7.64\cite{isajet}.
This version also includes our radiative neutralino decay calculation.

In Fig. \ref{fig1}, we show the $m_0\ vs.\ m_{1/2}$ plane for
$\tan\beta =10$, $A_0=0$ and $\mu >0$. The red-shaded regions
are excluded either by the presence of a stau LSP (upper left)
or a lack of REWSB (lower and right). The pink shaded region is 
excluded by LEP2 searches for charginos and a light Higgs scalar.
The upper and left regions indicate where $\tz_2$ two-body decays
occur: either to $\tell_1 \ell$ ($\ell$ denotes any of the charged
or neutral leptons) on the left, or to $\tz_1 Z$ or $\tz_1 h$ in the upper
plane. The remaining regions allowed by LEP2 have dominant three-body
$\tz_2$ decays. The green shaded region has 
$0.001<BF(\tz_2\to\tz_1\gamma )<0.005$ while the yellow has
$0.005<BF(\tz_2\to\tz_1\gamma )<0.01$. In these regions, 
where $m_{1/2}\alt 250$ GeV, we also have $m_{\tg}\alt 650$ GeV.
Then $\tg\tg$ cross sections exceed $\sim 2000$ fb at the CERN LHC, leading to
at least 20,000 SUSY events per 10 fb$^{-1}$ of integrated luminosity.
After suitable cuts are made on number of jets, 
jet $E_T$ and $\eslt$, a nearly pure sample of SUSY events will 
remain\cite{lhc}. These SUSY candidates can then
be scrutinized for the presence of hard, isolated photons,
and may allow a determination at some level of 
the neutralino radiative decay branching fraction. 

\FIGURE[t]{\epsfig{file=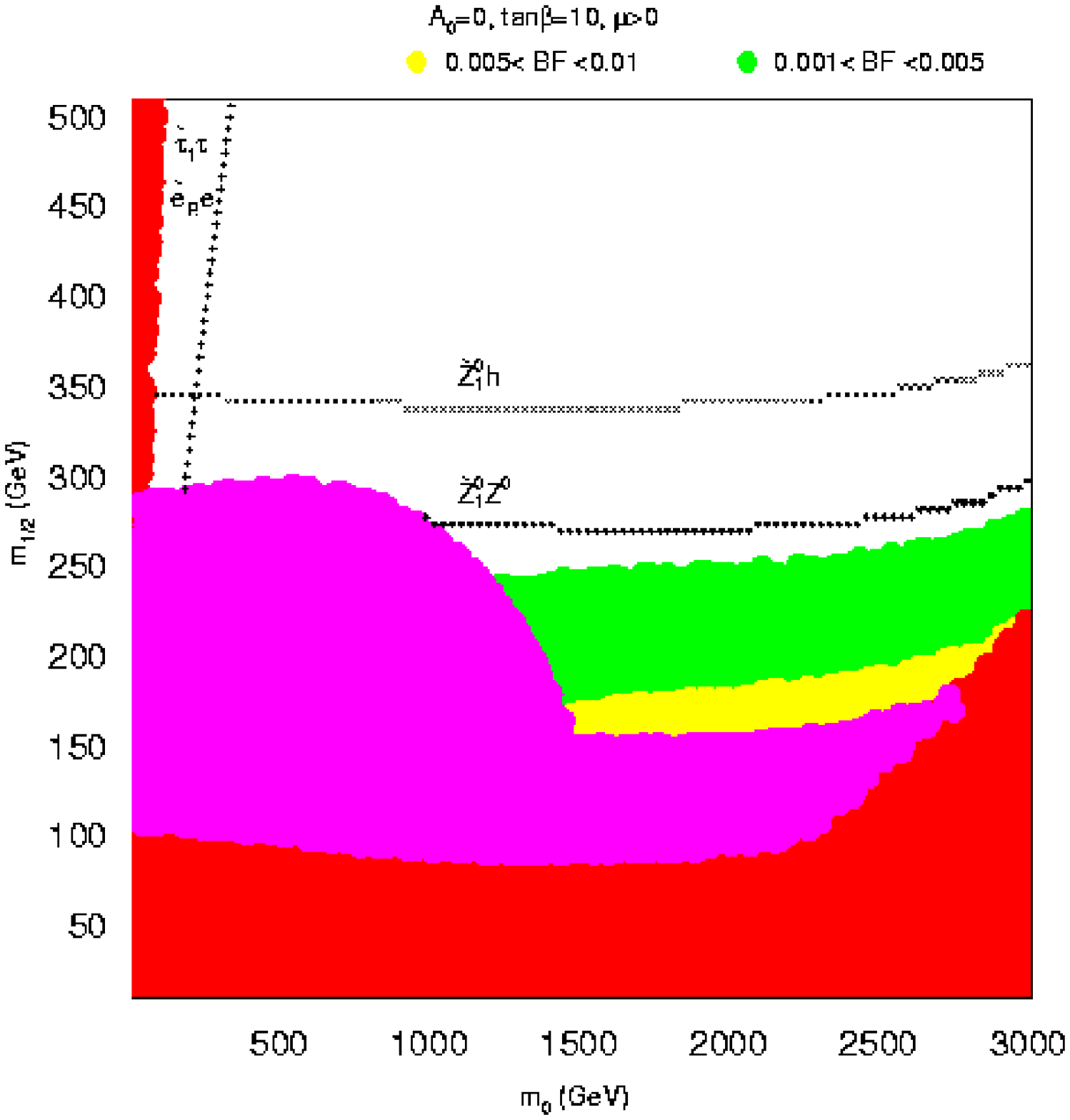,width=15cm} 
\caption{Regions of mSUGRA model parameter space with significant
$\tz_2\to\tz_1\gamma$ branching fraction, in the 
$m_0\ vs.\ m_{1/2}$ plane for $\tan\beta =10$, $A_0=0$ and $\mu >0$.
The red shaded regions are theoretically excluded while pink regions
are excluded by sparticle searches at LEP2. The green region
has $0.001<BF(\tz_2\to\tz_1\gamma )<0.005$, while the yellow
region has  $0.005<BF(\tz_2\to\tz_1\gamma )<0.01$.}
\label{fig1}}

We note that the regions of mSUGRA parameter space with the largest 
$\tz_2\to \tz_1\gamma$ branching fraction occur at rather
large $m_0$ and low $m_{1/2}$ values. The very largest allowed
$m_0$ areas correspond to the focus-point region of Feng, Matchev
and Moroi\cite{feng}, where fine-tuning may be low, but squark and slepton
masses are at the multi-TeV level. These heavy sfermion masses 
may be sufficient to
suppress many flavor-changing or $CP$ violating processes that could
arise due to the presence of off-diagonal flavor changing soft SUSY masses, 
or imaginary components of soft SUSY breaking terms. 

We also show the regions of $\tz_2\to\tz_1\gamma$ branching fraction
for large $\tan\beta =45$ in Fig. \ref{fig2}. 
Again, we see a significant part of the three-body 
$\tz_2$ decay region inhabited with possibly measurable branching fractions
for radiative neutralino decay. The region with the largest branching
fraction is adjacent to the right-side
region forbidden by lack of REWSB. In that region, the $\mu$ parameter
is becoming small, and the light neutralinos are becoming increasingly 
higgsino-like. In addition, sfermion masses are heavy, so we are in a region
of reduced dynamical suppression of three-body decay modes, also with some
kinematic suppression. 

In both the low and the high $\tan\beta$ cases, 
it is important to note that the $\tz_2\to\tz_1\gamma$ branching fraction 
reaches no greater than 1\%, so is never dominant. However, regions do occur
where the radiative decay should be detectable, and the branching fraction
should be measureable. Detailed simulations including detector effects for 
identifying isolated photons will be necessary to 
ascertain the precision with which such measurements can be made.

\FIGURE[t]{\epsfig{file=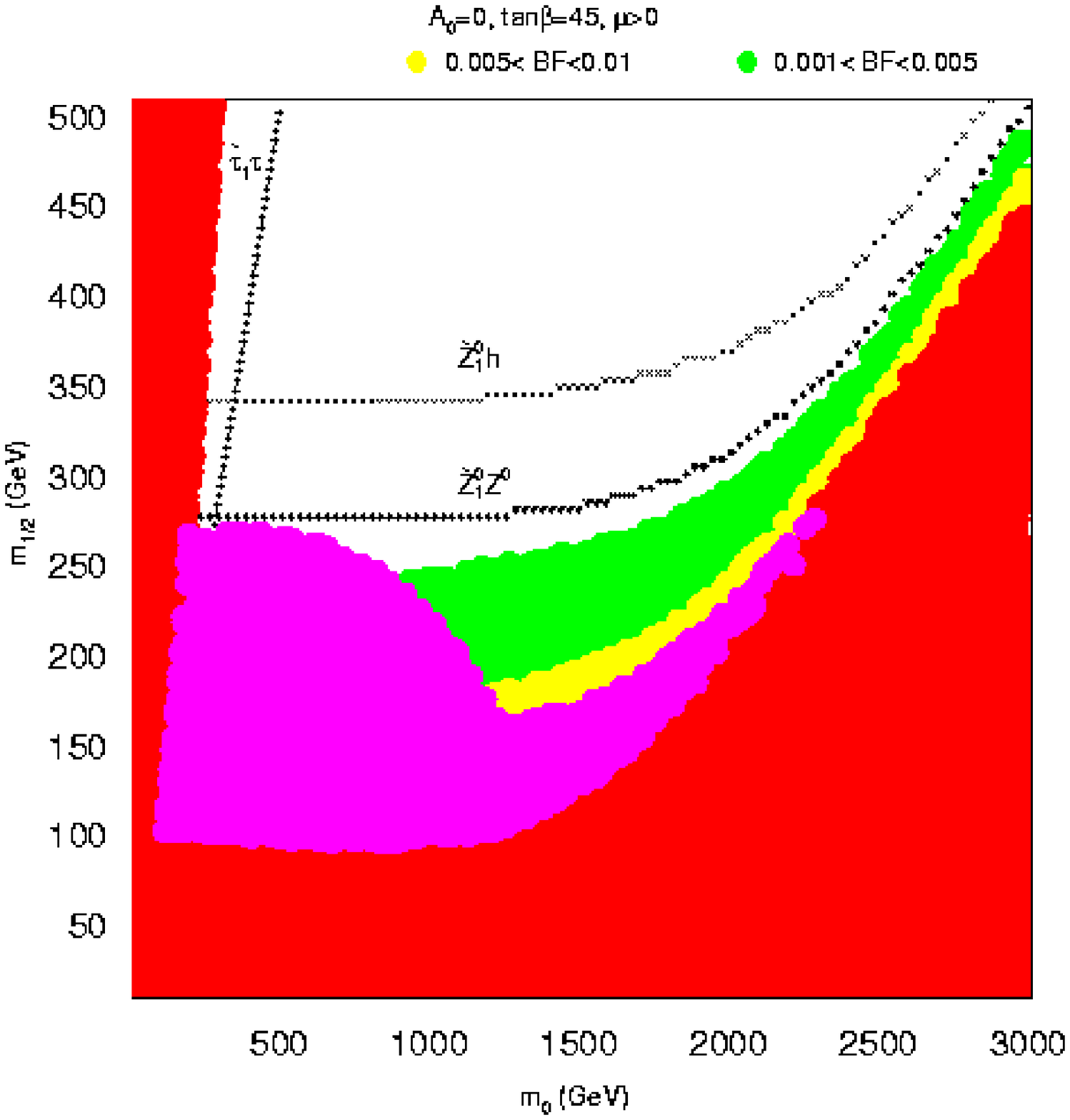,width=15cm} 
\caption{Same as Fig. 1, except $\tan\beta =45$.}
\label{fig2}}

\section{$\tz_2\to\tz_1\gamma$ branching fraction in the MSSM}
\label{sec:mssm}

It is apparent from the numerical studies of Ambrosanio and Mele\cite{mele} 
that
the greater freedom of parameter choices in the unconstrained MSSM can allow
for regions of parameter space with larger $\tz_2\to\tz_1\gamma$ branching
fractions than in the mSUGRA model. In this section, we present a scan over
unconstrained MSSM parameters. To make the analysis tractable, we do assume
all MSSM soft SUSY breaking parameters and Yukawa 
couplings to be purely real, and we take all sfermion masses to be 
degenerate, with no intergenerational mixing. 
For simplicity, we take $A_t=A_b=A_\tau$, and fix $m_{\tg}=1500$ GeV.
Our scan of MSSM 
parameter space is then over the following region:
\begin{eqnarray*}
-1500\ {\rm GeV}<&M_1&<1500\ {\rm GeV},\\
0 <&M_2&<1500\ {\rm GeV},\\
-1500\ {\rm GeV}<&\mu&<1500\ {\rm GeV},\\
-2000\ {\rm GeV}<&A_t&<2000\ {\rm GeV},\\
150\ {\rm GeV}<&m_A&<1500\ {\rm GeV},\\
2 <&\tan\beta &<50,\\
100\ {\rm GeV}<&m_{\tf}&<1500\ {\rm GeV},
\end{eqnarray*}
where we have the freedom to choose one of the gaugino masses, 
in this case $M_2$, to be real and positive.
We require $m_{\tw_1}>103.5$ GeV and $m_h>114.1$ GeV, according to LEP2.
The latter Higgs mass requirement is overly harsh in that the limit
on the lightest SUSY Higgs boson is somewhat lower, but this doesn't
alter our conclusions. We also require a mass gap 
$m_{\tz_2}-m_{\tz_1}>5$ GeV, so that the photon that appears in
scattering events from any of these models has a detectable energy.

Our results are presented in Figs. \ref{fig3} and \ref{fig4}, 
as plots of branching fraction ($BF$) versus model parameter. 
Fig. \ref{fig3}{\it a} and \ref{fig3}{\it b} show the branching fractions
versus the gaugino masses $M_1$ and $M_2$. We see immediately that in fact
there do exist models with substantial radiative decay branching fractions
in the range 10\%-100\%. For both gaugino masses, these large
branching fraction models populate the region with $|M_1|,\ M_2<1000$ GeV,
with the distribution of points being somewhat asymmetric for
$M_1$ being positive or negative. 
%If $|M_1|$ or $M_2$ becomes
%too large, either two-body decay modes of $\tz_2$ turn on, or
%else $\tz_2\to\tw_1 f\bar{f}'$ becomes large, thus supressing the 
%radiative decay branching fraction, unless $|M_1|\sim M_2$. 
% Tadas
If $M_1$ becomes large negative 
or $M_2$ becomes large, the branching fractions can remain large, but 
the mass gap $m_{\tz_2}-m_{\tz_1}$ decreases 
below 5 GeV. However, for large positive $M_1$ the mass gap stays large,
but in this case we do not get large branching fractions
because
either two-body decay modes of $\tz_2$ turn on, or else 
$\tz_2\to\tw_1 f\bar{f}'$ 
becomes large, thus supressing the radiative decay branching fraction.
% End
The models with $|M_1|,\ M_2$
near zero are usually excluded by constraints from LEP on $Z\to\tz_i\tz_j$
decay or from LEP2 on the chargino mass.

\FIGURE[t]{\epsfig{file=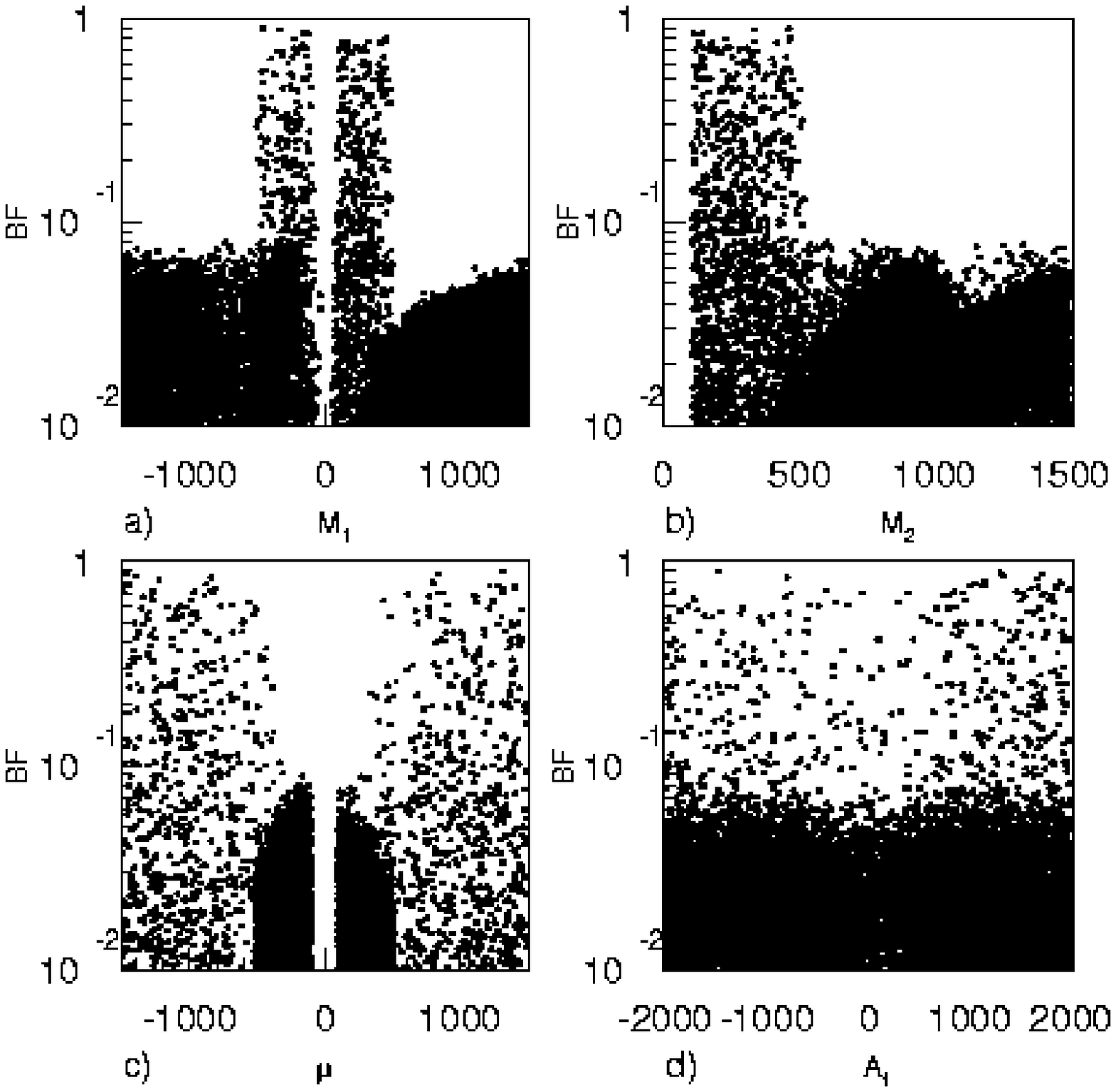,width=15cm} 
\caption{Plot of $BF(\tz_2\to\tz_1\gamma )$ versus the parameter
{\it a}) $M_1$, {\it b}) $M_2$, {\it c}) $\mu$ and {\it d}) $A_t$ in the 
MSSM model.} \label{fig3}}

Fig. \ref{fig3}{\it c} shows the $BF$ versus the $\mu$ parameter.
Again, small values of $|\mu|$ are excluded by LEP and LEP2 constraints.
However, we see that really large radiative decay branching fractions 
only occur for very large values of $|\mu |\agt 300-400$ GeV.
%If $|\mu |\ll |M_1|,\ M_2$, then the two lightest neutralinos
%are both higgsino-like, and $\tz_2\to\tz_1 f\bar{f}$ decays via $Z^*$
%are large, and suppress the radiative decay branching fraction.
%
The suppresion of lower values of $|\mu|$ comes from the mass gap 
requirement.
Fig. \ref{fig3}{\it d} shows the branching fraction versus the
parameter $A_t$. Here, large values of branching fraction are possible for all
values of $A_t$, although the very largest seem to favor large values
of $|A_t|$.

In Fig. \ref{fig4} we show the $BF$ versus the parameters
$m_A$, $\tan\beta$ and $m_{\tf}$ in frames {\it a}, {\it b}
and {\it c}. We see that large branching fractions are possible
for all values of pseudoscalar mass $m_A$, and for all 
values of $\tan\beta$. Large values of BF have some preference for
$m_{\tf}\agt 300-400$ GeV, so that two-body decays $\tz_2\to f\tf$
are closed, and three-body decays $\tz_2\to\tz_1 f\bar{f}$ through
an intermediate $\tf$ have some suppression.

\FIGURE[t]{\epsfig{file=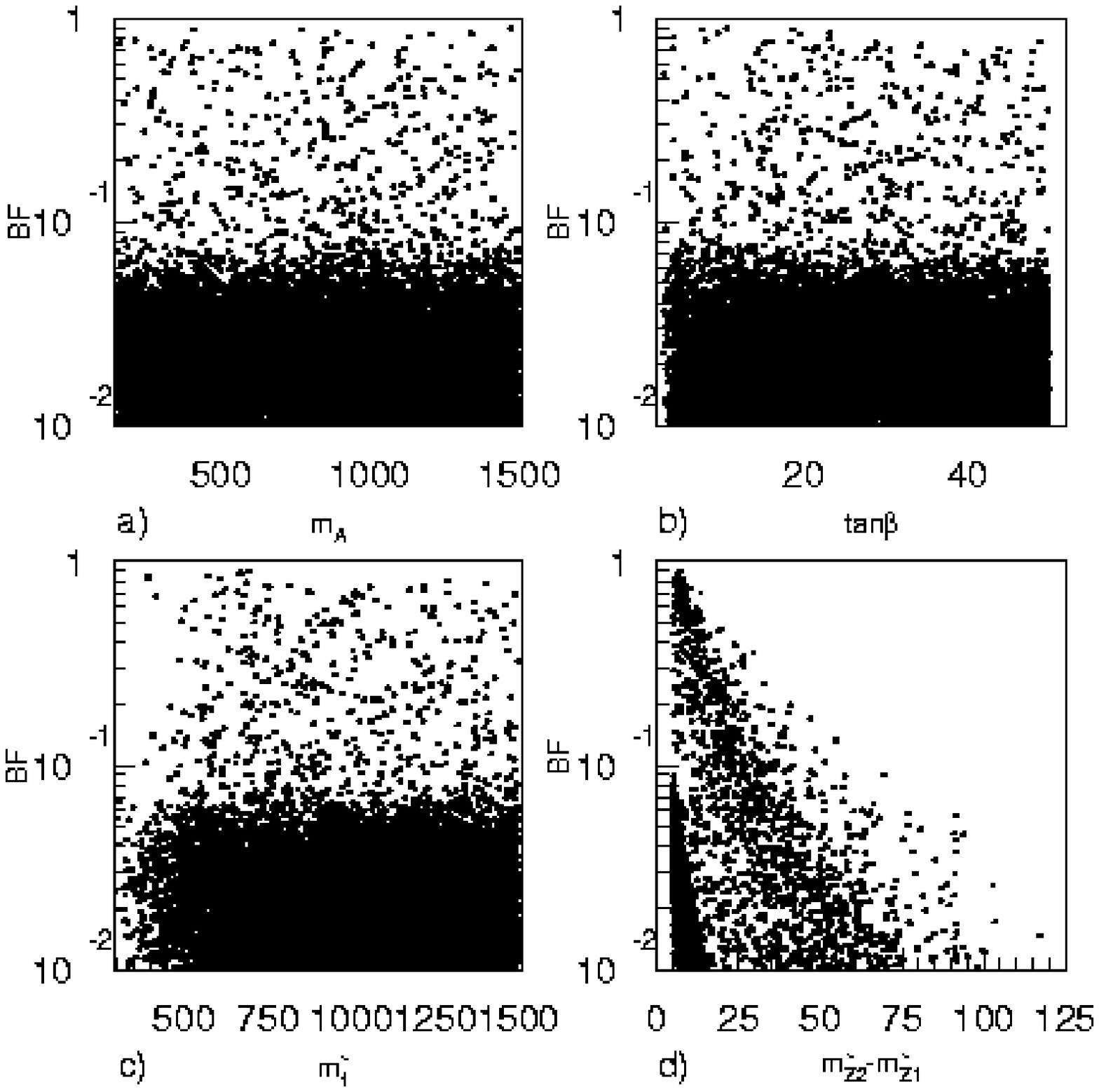,width=15cm} 
\caption{Plot of $BF(\tz_2\to\tz_1\gamma )$ versus the parameter
{\it a}) $m_A$, {\it b}) $\tan\beta$, {\it c}) $m_{\tf}$ and
{\it d}) $m_{\tz_2}-m_{\tz_1}$ in the MSSM model.}
\label{fig4}}

In the last frame, Fig. \ref{fig4}{\it d}, we show the branching fraction 
not against an input parameter, but against the neutralino mass
gap, $m_{\tz_2}-m_{\tz_1}$. Here it is clear that the largest values
of $BF$ prefer the smallest mass gaps, where there exists 
at least some amount of kinematical
suppression of three-body $\tz_2$ decays. The large branching fractions
have a cut-off around 90-100 GeV, where $\tz_2\to\tz_1 Z$ decays
become possible.

To gain more understanding of the parameter space where
radiative decays are large, we plot in Fig. \ref{fig5} 
only models with $BF>10\%$
(black dots) or $>1\%$ (red dots), in correlated parameter planes.
Again, we require $m_{\tz_2}-m_{\tz_1}>5$ GeV, and all LEP and LEP2
constraints to be satisfied. From Fig. \ref{fig5}{\it a} in the
$M_1\ vs.\ M_2$ plane, we see that models with the largest $BF$
require $|M_1|\sim M_2$, a feature noted by Ambrosanio
and Mele in their analysis\cite{mele}. The region with
$M_1\sim M_2$ turns out to be quite narrow, while the region 
with $M_1\sim -M_2$ has slightly more breadth to it. This latter point is 
even more apparent when we relax to the condition $BF>1\%$.

In Fig. \ref{fig5}{\it b} and \ref{fig5}{\it c}, we show models
in the $M_1\ vs.\ \mu$ and $M_2\ vs.\ \mu$ planes. For models
with $BF>10\%$, evidently $|M_1|<|\mu |$ and also $M_2 <|\mu |$.
If we relax to models with $BF>1\%$, then these requirements no longer hold.
In frame \ref{fig5}{\it d}, we show models in the $M_2/M_1\ vs.\ \mu$
plane. Models with $BF>10\%$ clearly favor $M_2/M_1\sim 1$, with either
sign of $\mu$ being equally preferable. 
%Again, there exists a broader
%region where models with large branching fractions are possible
%when $M_2/M_1 <0$, so that the gaugino masses have opposite signs.
If we allow models with $BF>1\%$, then the $M_1\sim \pm M_2$ 
restriction can be relaxed, and rather large ratios of $M_2/M_1$ are allowed,
especially if $|\mu |$ is small.

\FIGURE[t]{\epsfig{file=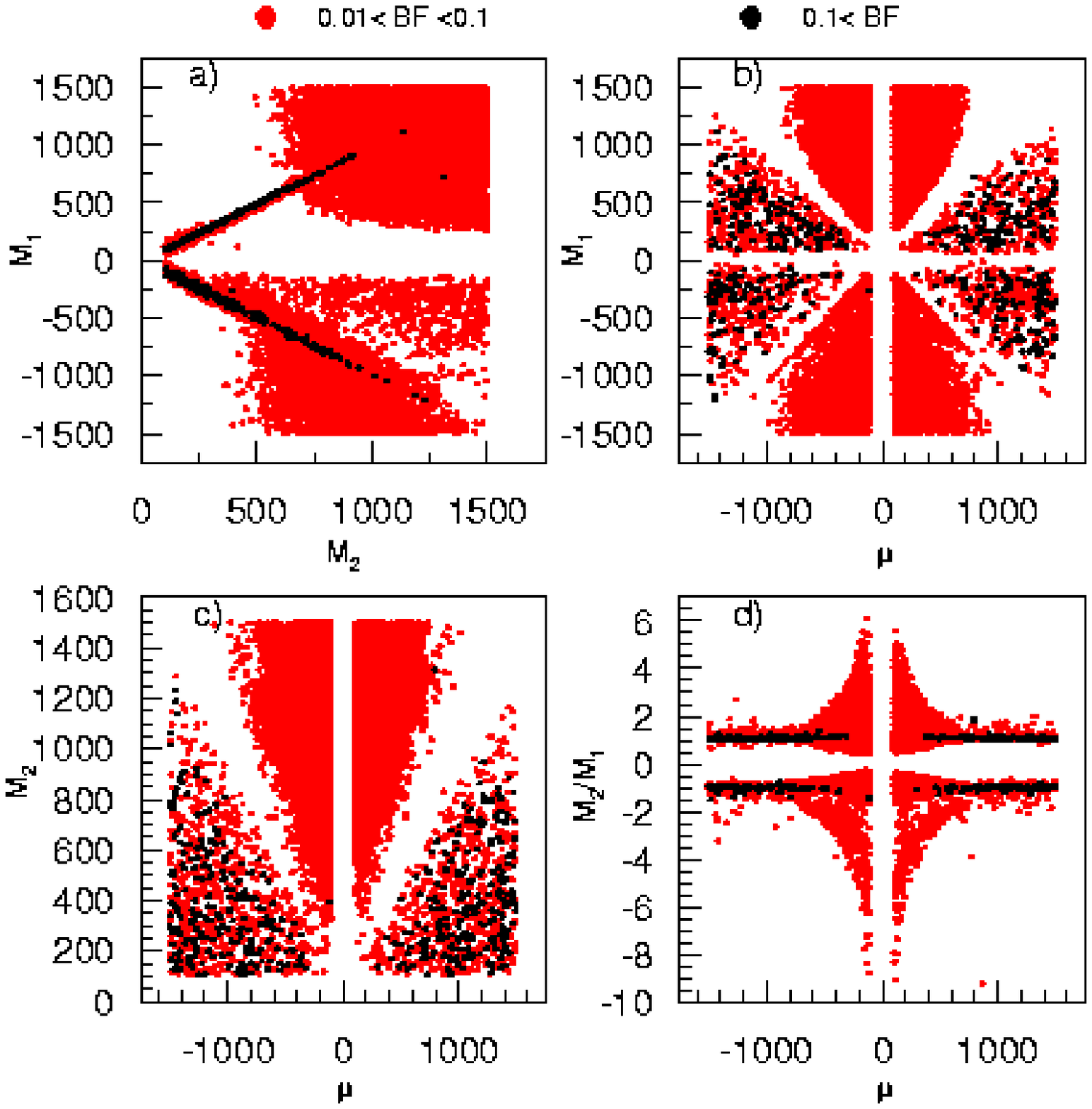,width=15cm} 
\caption{Plot of MSSM model points with $BF(\tz_2\to\tz_1\gamma )>0.1$
and $m_{\tz_2}-m_{\tz_1}>5$ GeV in {\it a}) $M_1\ vs.\ M_2$
space, {\it b}) $M_1\ vs.\ \mu$ space, {\it c}) $M_2\ vs.\ \mu$
space and {\it d}) $M_2/M_1\ vs.\ \mu$ space.}
\label{fig5}}

An important element of our analysis is the size of the $\tz_2 -\tz_1$
mass gap, so there is sufficient photon energy for observations.
In Fig. \ref{fig6}, we show the mass gap $m_{\tz_2}-m_{\tz_1}$ versus
the parameters $\mu$, $M_1$, $M_2$ and $\tan\beta$ in frames
{\it a}-{\it d}, respectively. From Fig. \ref{fig6}{\it a}, we see
that models with $BF>10\%$ prefer large absolute values of the
$\mu$ parameter, but for these models the mass gap is bounded 
typically by 40-60 GeV. By relaxing to models with $BF>1\%$, we pick up models
with mass gaps ranging to 100 GeV. However, for models 
with small values of $\mu$, where higgsino$\to$ higgsino transitions
are possible, the mass gap is very small ($\alt 20$ GeV), and 
these cases will be more difficult to observe. 
Figs. \ref{fig6}{\it b} and \ref{fig6}{\it c}
show that the mass gap for models with $BF>10\%$ is typically less than
30-40 GeV for $|M_1|,\ M_2$ ranging up to 1000 GeV. The models with
$BF>1\%$ and $|M_1|,\ M_2> 1000$ GeV typically 
have a small mass gap, less than 10-15 GeV. 
Finally, in Fig. \ref{fig6}{\it d}, the mass gap versus
$\tan\beta$ shows no major correlations, and significant mass gaps can 
be achieved for any $\tan\beta$ values.  

\FIGURE[t]{\epsfig{file=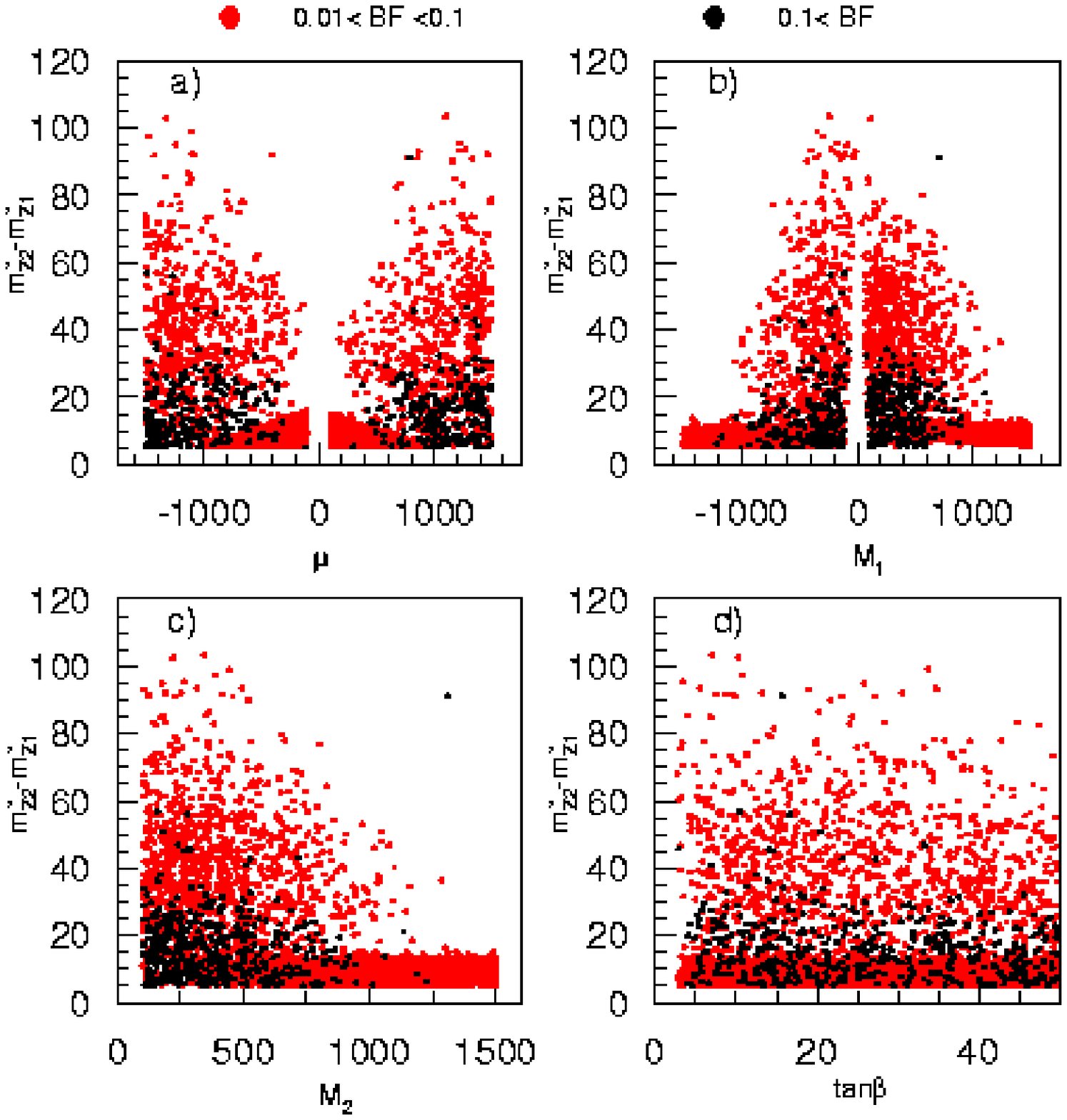,width=15cm} 
\caption{Plot of the mass gap $m_{\tz_2}-m_{\tz_1}$ versus
{\it a}) $\mu$, {\it b}) $M_1$ and {\it c}) $M_2$ for MSSM model points 
with $BF(\tz_2\to\tz_1\gamma )>0.1$
and $m_{\tz_2}-m_{\tz_1}>5$ GeV. In {\it d}), we plot the
mass gap $m_{\tz_2}-m_{\tz_1}$ versus the parameter $\tan\beta$
in the MSSM.}
\label{fig6}}

To conclude this section, we summarize our findings by noting that the 
largest neutralino radiative decay branching fractions are found in regions
of parameter space with
\begin{eqnarray*}
|M_1|,\ M_2 \sim 100-1000\ {\rm GeV},\\
|\mu |>|M_1|,\ M_2 \ \ {\rm and}\\
|M_1|\sim M_2 .
\end{eqnarray*}
%with some preference being give to $M_1\sim -M_2$, so that the gaugino masses
%have opposite signs.

\section{$\tz_2\to\tz_1\gamma$ branching fraction in a SUGRA model
with non-universal gaugino masses}
\label{sec:nusug}

So far, our findings are that in the mSUGRA model the radiative decay
$BF$ never exceeds 1\%, and is largest in regions with large $m_0$
such as the focus point region. For the MSSM case, we found 
distinct regions of parameter space where the radiative branching fraction 
can range up to 10\%-100\%, although these conditions are never realized 
in the mSUGRA model.
In this section, we note that there {\it do} exist models where the above 
MSSM parameter space restrictions are realized, namely in SUGRA models
with non-universal gaugino masses.

Since supergravity
is not a renormalizable theory, in general we may
expect a non-trivial gauge kinetic function.
Expanding the gauge kinetic function as
\begin{equation}
f_{AB} = \delta_{AB} + \hhat_{AB}/M_P + \ldots,
\end{equation}
where the fields
$\hhat_{AB}$ transform as left handed chiral superfields
under supersymmetry transformations, and as
the symmetric product of two adjoints under gauge symmetries, we parametrize
the lowest order contribution to gaugino masses by,
\begin{equation}
\CL \supset -{1\over 4}\int d^2\th_L \frac{\hhat_{AB}}{M_P}
\overline{\hat{W}_A^c}\hat{W}_B\supset \frac{\langle F_h\rangle_{AB}}{M_P}
\lambda_A\lambda_B +\ldots
\end{equation}
where the $\hat{W}_A$ are curl superfields, the $\lambda_A$ 
are the gaugino fields,
and $F_{hAB}$ are the auxillary field components of the $\hhat_{AB}$ that
acquire a SUSY breaking $vev$.

If the fields $F_{h}$ which break supersymmetry are gauge singlets,
universal gaugino masses result.  However, in principle, the chiral
superfield which communicates supersymmetry breaking to the gaugino
fields can lie in any representation in the symmetric product of two
adjoints, and so can lead to gaugino mass terms
that (spontaneously) break the underlying gauge symmetry. 

In the context of $SU(5)$
grand unification, $F_{h}$ belongs to an $SU(5)$ irreducible
representation which appears in the symmetric product of two adjoints:
\begin{equation}
({\bf 24}{\bf \times}
 {\bf 24})_{\rm symmetric}={\bf 1}\oplus {\bf 24} \oplus {\bf 75}
 \oplus {\bf 200}\,,
\end{equation}

where only $\bf 1$ yields universal masses.  
The relations amongst the
various $GUT$ scale gaugino masses have been worked out by 
Anderson\cite{anderson}.
The relative $GUT$ scale $SU(3)$, $SU(2)$ and
$U(1)$ gaugino masses $M_3$, $M_2$ and $M_1$ are listed in
Table~\ref{tab:and} along with the approximate masses after RGE evolution
to $Q\sim M_Z$.  
These scenarios represent the predictive subset
of the more general (and less predictive) case of an arbitrary superposition
of these
representations.
The model parameters may be chosen to be, 
\begin{equation}
m_0,\ M_3^0,\ A_0,\ \tan\beta\ {\rm and}\ sign(\mu ),
\end{equation}
where $M_i^0$ is the $SU(i)$ gaugino mass at scale $Q=M_{GUT}$.  $M_2^0$
and $M_1^0$ can then be calculated in terms of $M_3^0$ according to
Table \ref{tab:and}, and the weak scale SUSY particle spectrum
can again be calculated using ISAJET v7.64.
\begin{table}
\begin{center}
\begin{small}
\begin{tabular}{|c|ccc|ccc|}
\hline
\ & \multicolumn{3}{c|} {$M_{GUT}$} & \multicolumn{3}{c|}{$M_Z$} \cr
$F_{h}$
& $M_3$ & $M_2$ & $M_1$
& $M_3$ & $M_2$ & $M_1$ \cr
\hline
${\bf 1}$   & $1$ &$\;\; 1$  &$\;\;1$   & $\sim \;6$ & $\sim \;\;2$ &
$\sim \;\;1$ \cr
${\bf 24}$  & $2$ &$-3$      & $-1$  & $\sim 12$ & $\sim -6$ &
$\sim -1$ \cr
 ${\bf 75}$  & $1$ & $\;\;3$  &$-5$      & $\sim \;6$ & $\sim \;\;6$ &
$\sim -5$ \cr
${\bf 200}$ & $1$ & $\;\; 2$ & $\;10$   & $\sim \;6$ & $\sim \;\;4$ &
$\sim \;10$ \cr
\hline
\end{tabular}
\end{small}
\smallskip
\caption{Relative gaugino masses at $M_{GUT}$ and $M_Z$
in the four possible $F_{h}$ irreducible representations.}
\label{tab:and}
\end{center}
\end{table}

We note that the {\bf 75} model leads to weak scale gaugino masses
in the ratio $M_2:M_1 \sim 6:-5$, which is approaching the conditions
required in Sec. \ref{sec:mssm} for models with large radiative
branching fractions. We adopt the GUT scale gaugino mass boundary
conditions of the {\bf 75} model, and explore the model parameter
space for large radiative decay branching fractions.

Our first results are shown in Fig. \ref{fig7} in the $M_3^0\ vs.\ m_0$
plane, for $\tan\beta =5$, $A_0=0$ and $\mu >0$.
Again, the red shaded regions are excluded theoretically, while the 
pink regions are excluded by constraints from LEP2. 
The region with mass gap $m_{\tz_2}-m_{\tz_1}>5$ GeV is indicated between
the dotted contours.
The substantial
green region has branching fractions $BF:\ 1\% - 10\%$, while the yellow
region has branching fractions in the range $BF:\ 10\% -50\%$.
Almost all the large branching fraction region has mass gap
$m_{\tz_2}-m_{\tz_1}<5$ GeV.
The value of $|\mu |$ in the yellow region is generally much smaller
than $|M_1|$ and $M_2$, so that $\tz_2\to\tz_1\gamma$ consists of a 
higgsino $\to$ higgsino transition, with a rather small mass gap
between $\tz_2$ and $\tz_1$; there is significant kinematic suppression of 
three-body decay rates. In the lower right unshaded regions, the value
of $\mu$ is comparable or larger than $|M_1|$ and $M_2$, 
and the $\tz_2$-$\tz_1$
mass gap is usually larger, so three-body decay rates increase, suppressing
the radiative decay branching fraction.
An exception occurs near $m_0\sim 3000$ GeV and $M_3^0\sim 300$ GeV
where the light neutralinos are transitioning between higgsino-like and
gaugino-like. In this area, the mass gap drops to very small values of order
1 GeV, and the radiative decay is momentarily enhanced. 

\FIGURE[t]{\epsfig{file=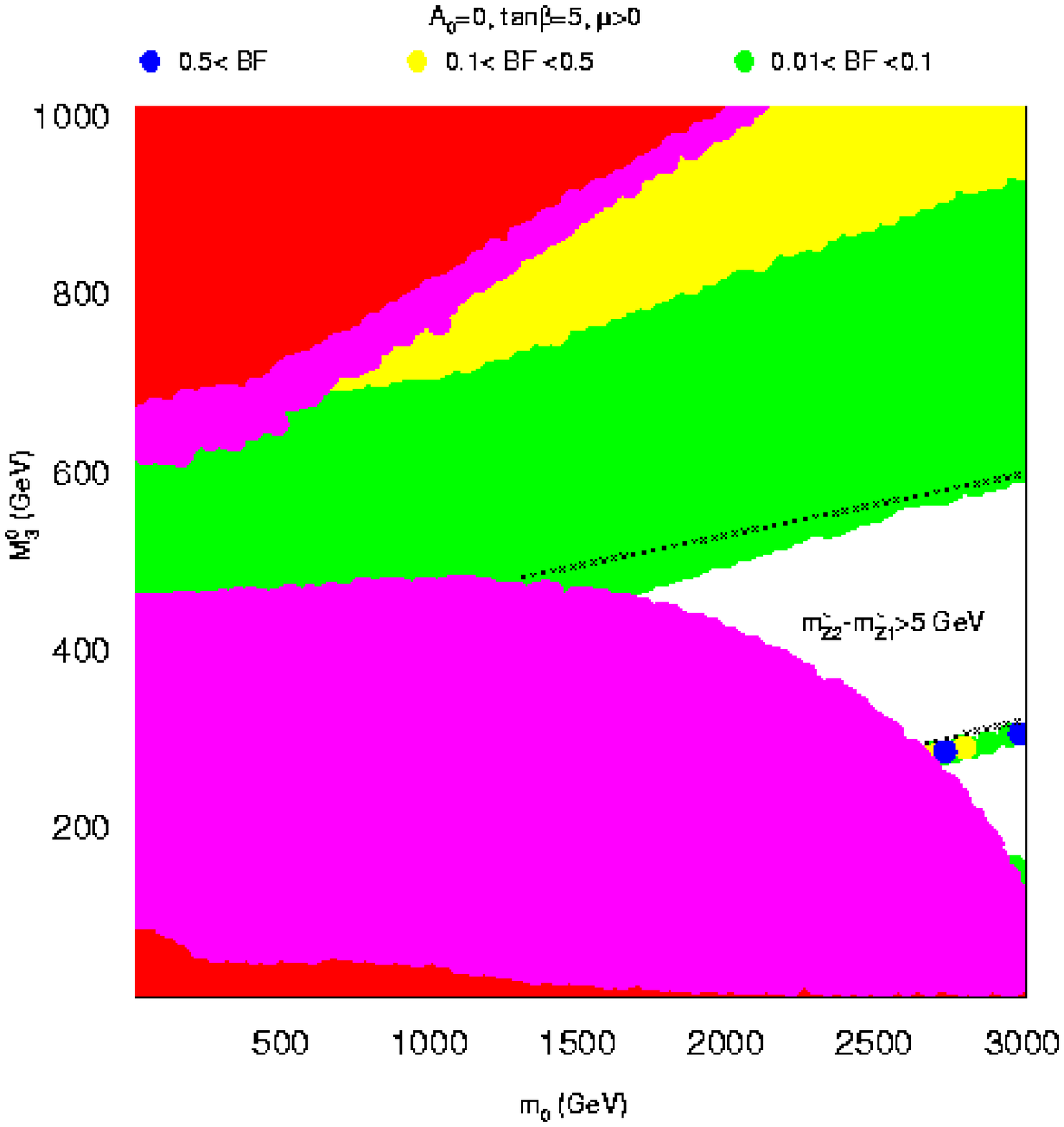,width=15cm} 
\caption{Regions of the {\bf 75} SUGRA model parameter space with significant
$\tz_2\to\tz_1\gamma$ branching fraction, in the 
$m_0\ vs.\ m_{1/2}$ plane for $\tan\beta =5$, $A_0=0$ and $\mu >0$.
The red shaded regions are theoretically excluded while pink regions
are excluded by sparticle searches at LEP2. The green region
has $0.01<BF(\tz_2\to\tz_1\gamma )<0.1$, while the yellow
region has  $0.1<BF(\tz_2\to\tz_1\gamma )<0.5$.}
\label{fig7}}
\newpage

In Fig. \ref{fig8}, we show again the {\bf 75} model, but this time with
$\tan\beta =10$. As $\tan\beta$ is increased beyond 5, the allowed 
parameter space rapidly decreases. In this case, only a small
region in the center of the plot gives a viable SUSY spectrum satsifying
all constraints. In this case, the yellow region represents 
$1\%<BF<5\% $ (lower values than in Fig. \ref{fig7}). The larger
$\tan\beta$ value allows for a larger mass gap between
$\tz_2$ and $\tz_1$ for small $\mu$ values   compared to Fig. \ref{fig7}
($m_{\tz_2}-m_{\tz_1}>5$ GeV throughout the $\tan\beta =10$ plane).
Consequently, three-body decays suffer less kinematical suppression,
and the radiative branching fraction is lower.

\FIGURE[t]{\epsfig{file=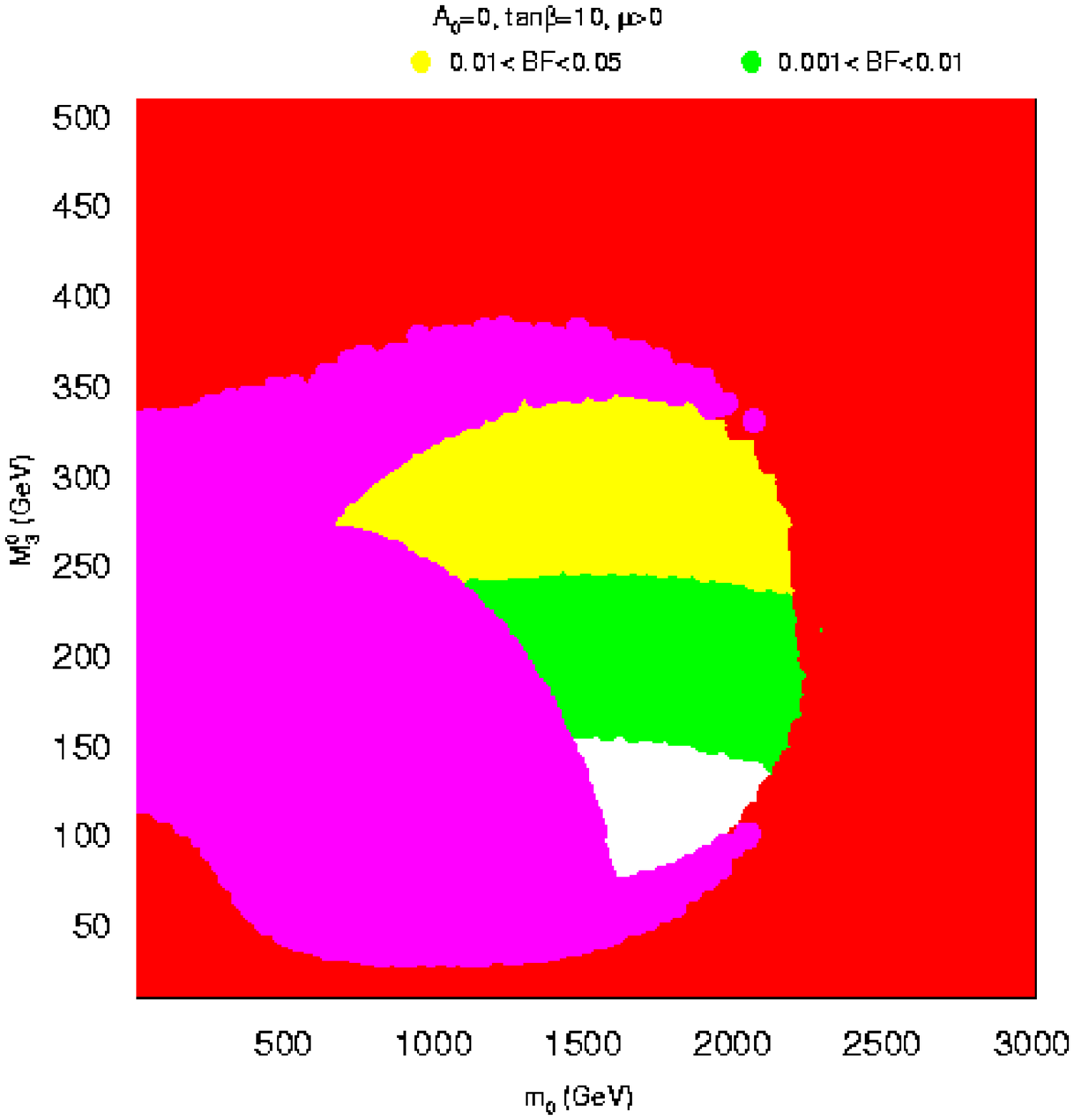,width=15cm} 
\caption{Same as Fig. 6, except $\tan\beta =10$, and now
the green region
has $0.001<BF(\tz_2\to\tz_1\gamma )<0.01$, while the yellow
region has  $0.01<BF(\tz_2\to\tz_1\gamma )<0.05$.}
\label{fig8}}
\newpage

We have also examined briefly two  cases of ``more minimal'' supersymmetric 
models\cite{nelson}. 
In these models, also known as inverted mass hierarchy (IMH) models,
there can exist non-universal scalar masses, leading to multi-TeV
first and second generation scalars (to suppress flavor and CP violating
SUSY processes), while third generation scalars remain
in the sub-TeV regime (as required by naturalness constraints).
The two models examined were the radiative inverted hierarchy
model (RIMH)\cite{rimh} and the GUT scale IMH model (GSIMH)\cite{gsimh}. 
In both these models,
usually two-body $\tz_2$ decays are allowed. In the RIMH model, 
for some parameter regions the $\mu$ parameter can become very small.
In that case, results are similar to the SUGRA model focus point
region already presented
with branching fractions near the $\sim 1\%$ level. 

%Alternatively,
%one can examine these models in an MSSM context, taking non-degenerate
%squark and slepton masses. We did not pursue an analysis with this additional
%parameter freedom. The analysis might make sense under the assumption of 
%gaugino mass unification, where the effects of the non-degenerate sfermions
%wouldn't be masked by the enhanced decays by non-universal gaugino masses.  

\section{Conclusions}
\label{sec:conclude}

We have re-calculated the decay rate for radiative neutralino decay
within the MSSM. We have also made detailed scans of parameter space in the 
mSUGRA model, the MSSM and a SUGRA model with non-universal gaugino masses.
In the mSUGRA model, radiative neutralino branching fractions 
for $\tz_2$ can reach up to 1\%. While this rate comprises only a small
fraction of the $\tz_2$ decay modes, it may be sufficiently large 
to be measureable at collider experiments at the CERN LHC, where many
$\tg$ and $\tq$ events can be produced, and $\tz_2$ will occur
in cascade decays. The rate is largest in the focus point region
of parameter space. If the branching fraction is observed, then it may aid
in the determination of underlying model parameters. 

In the MSSM, the $\tz_2\to\tz_1\gamma$ decay rate can reach nearly
100\%. In this case, the gaugino masses
$|M_1|,\ M_2\alt 1000$ GeV, with $|M_1|,\ M_2<|\mu |$ and
$|M_1|\sim M_2$. The branching fraction can be large for all values of 
$\tan\beta$. 
%There is some preference for the gaugino masses
%$M_1$ and $M_2$ to be of opposite sign. 

The conditions on parameters  of the MSSM for a large radiative
neutralino branching fraction turn out to be realized in a SUGRA
model with non-universal gaugino masses, where the hidden sector
field $\hhat_{AB}$ transforms as a 75 dimensional
symmetric product of two adjoints of $SU(5)$. At low $\tan\beta \sim 5$,
in this case the branching fraction can reach up to 50\%, although
the mass gap $m_{\tz_2}-m_{\tz_1}$ tends to be rather small - of order
a few GeV. 
For higher $\tan\beta$, the parameter space in the {\bf 75} model rapidly 
diminishes. However, scans made for the $\tan\beta =10$ case show the 
radiative branching fraction reaching up to $\sim 5\%$.

Our final conclusion is that once a significant sample of SUSY particle 
candidate events is discovered at collider experiments, then that sample 
should be scrutinized for the presence of additional isolated
photons, beyond the level expected from bremsstrahlung, 
misidentified jets, etc. If the events contain in addition isolated photons,
then it may be possible to test whether they come from radiative
neutralino decay, and to in fact measure that particular
branching fraction. If the $\tz_2\to\tz_1\gamma$ branching
fraction is in fact measured, then its value should aid in the attempt to
extract the values of underlying model parameters.

\section*{Appendix}

In this appendix, we present our
formulae for the radiative neutralino decay rate.
The complete radiative decay width of $\tilde Z_j$ was 
calculated by Haber and Wyler\cite{haber}. We organize our calculation
in a similar fashion to theirs, working in the non-linear $R_\xi$ gauge. 
We use the Lagrangian conventions of Tata\cite{reviews}, which is 
useful for inclusion into ISAJET.
The relevant Feynman diagrams
labelled {\it a}-{\it p} are presented in Ref. \cite{haber}, 
and will not be repeated here.
We have checked that our formula reproduces analytically that of 
Haber and Wyler after converting conventions.
%(after several typos are corrected in their paper).

In addition, we have compared our numerical results for
various diagram subsets with a calculation by Ambrosanio and Mele which is
embedded in the SUSYGEN 3.00/36 program\cite{susygen}. 
Again, we find excellent agreement,
except for the case of sfermion loop diagrams, where mixing effects 
cause our result to differ from those of SUSYGEN. 
If we neglect mixing effects, 
then we agree with the SUSYGEN result.

The radiative decay width is given by
\begin{equation}
\Gamma(\tilde Z_j\rightarrow\tilde Z_i\gamma)=
\frac{g^2_{\tilde Z_j\tilde Z_i\gamma}}{8\pi}
\left (\frac{m^2_{\tilde Z_j}-m^2_{\tilde Z_i}}{m_{\tilde Z_i}}\right )^3.
\end{equation}
Here
\begin{equation}
g_{\tilde Z_j\tilde Z_i\gamma}=g_{abcd}+g_{efgh}+g_{ijkl}+g_{mnop},
\end{equation}
with the separate contributions given by the following:
\begin{equation}
g_{abcd}=\sum_{sfermions}
\frac{eq_f}{(4\pi)^2}C_f\left (m_{\tilde Z_i}C^{abcd}_i K
+m_{\tilde Z_j}C^{abcd}_j (I^2-K)+m_fC^{abcd}_0 I\right ).
\end{equation}
Here $C_f=1$ for leptons and $C_f=3$ for quarks.
\begin{eqnarray*}
C^{abcd}_i&=&(-i)^{\theta_i}(i)^{\theta_j}\left (\alpha^{(i)}_k
\alpha^{(j)}_k-\beta^{(i)}_k \beta^{(j)}_k\right ),\\
C^{abcd}_j&=&-(i)^{\theta_i}(-i)^{\theta_j}\left (\alpha^{(i)}_k
\alpha^{(j)}_k-\beta^{(i)}_k \beta^{(j)}_k\right ),\\
C^{abcd}_0&=&(i)^{\theta_i+\theta_j}\left (\beta^{(i)}_k
\alpha^{(j)}_k-\alpha^{(i)}_k \beta^{(j)}_k\right ),\\
\alpha^{(i)}_{\tilde f_1}&=&A^{(i)}_f\cos\theta_f-f_f v^{(i)}_a\sin\theta_f,\\
\alpha^{(i)}_{\tilde f_2}&=&A^{(i)}_f\sin\theta_f+f_f v^{(i)}_a\cos\theta_f,\\
\beta^{(i)}_{\tilde f_1}&=&-B^{(i)}_f\sin\theta_f+f_f v^{(i)}_a\cos\theta_f,\\
\beta^{(i)}_{\tilde f_2}&=&B^{(i)}_f\cos\theta_f+f_f v^{(i)}_a\sin\theta_f,\\
A^{(i)}_u&=&\frac{1}{\sqrt2}\left (gv^{(i)}_3+\frac{g^\prime}{3}v^{(i)}_4\right ),\\
A^{(i)}_d&=&\frac{1}{\sqrt2}\left (-gv^{(i)}_3+\frac{g^\prime}{3}v^{(i)}_4
\right ),\\
A^{(i)}_l&=&-\frac{1}{\sqrt2}\left (gv^{(i)}_3+g^\prime v^{(i)}_4 \right ),\\
B^{(i)}_u&=&-\frac{4}{3\sqrt2}g^\prime v^{(i)}_4,\\
B^{(i)}_d&=&\frac{2}{3\sqrt2}g^\prime v^{(i)}_4,\\
B^{(i)}_l&=&\sqrt2 g^\prime v^{(i)}_4.\\
\end{eqnarray*}
For the neutralino mixing element $v_a^{(i)}$, 
$a=1$ if $f$ is an up type quark while $a=2$ for leptons and down type quarks.

The $g$ and $g'$ are the $SU(2)$ and $U(1)_Y$ gauge couplings, while
the $v_i^{(j)}$ are neutralino mixing elements, as defined
by Tata\cite{reviews}. The $f_f$ is the Yukawa coupling for fermion $f$,
and $\theta_f$ is the sfermion mixing angle. The integrals
$I$, $I^2$, $J$ and $K$ are defined in Haber and Wyler\cite{haber}.

Also,
\begin{equation}
g_{efgh}=\frac{e}{8\pi^2}\left (m_{\tilde Z_i}C^{efgh}_i(J-K)
-m_{\tilde Z_j}C^{efgh}_j (I^2-J-K)-2m_{\tw_k}C^{efgh}_0 J\right ),
\end{equation}
where
\begin{eqnarray*}
C^{efgh}_i&=&(-i)^{\theta_i}(i)^{\theta_j}\left (D^{(i)}_k
D^{(j)}_k-E^{(i)}_k E^{(j)}_k\right ),\\
C^{efgh}_j&=&-(i)^{\theta_i}(-i)^{\theta_j}\left (D^{(i)}_k
D^{(j)}_k-E^{(i)}_k E^{(j)}_k\right ),\\
C^{efgh}_0&=&(i)^{\theta_i+\theta_j}\left (E^{(i)}_k
D^{(j)}_k-D^{(i)}_k E^{(j)}_k\right ),\\
D^{(i)}_1&=&g(-1)^{\theta_{\tilde W_1}}\left (\frac{\cos\gamma_R}{\sqrt2}
v^{(i)}_1+\sin\gamma_R v^{(i)}_3\right ),\\
D^{(i)}_2&=&g(-1)^{\theta_{\tilde W_2}}\theta_y\left (-\frac{\sin\gamma_R}
{\sqrt2}v^{(i)}_1+\cos\gamma_R v^{(i)}_3\right ),\\
E^{(i)}_1&=&g\left (-\frac{\cos\gamma_L}
{\sqrt2}v^{(i)}_2+\sin\gamma_L v^{(i)}_3\right ),\\
E^{(i)}_2&=&g\theta_x \left (\frac{\sin\gamma_L}{\sqrt2}
v^{(i)}_2+\cos\gamma_L v^{(i)}_3\right ).\\
\end{eqnarray*}
Furthermore,
\begin{equation}
g_{ijkl}=-\frac{e}{(4\pi)^2}\left (m_{\tilde Z_i}C^{ijkl}_i K
+m_{\tilde Z_j}C^{ijkl}_j (I^2-K)+m_{\tw_k}C^{ijkl}_0 I\right ).
\end{equation}
\begin{eqnarray*}
C^{ijkl}_i&=&(-i)^{\theta_i}(i)^{\theta_j}\left (\sin^2\beta A^{(i)}_{k2}
A^{(j)}_{k2}-\cos^2\beta A^{(i)}_{k1}A^{(j)}_{k1}\right ),\\
C^{ijkl}_j&=&-(i)^{\theta_i}(-i)^{\theta_j}\left (\sin^2\beta A^{(i)}_{k2}
A^{(j)}_{k2}-\cos^2\beta A^{(i)}_{k1}A^{(j)}_{k1}\right ),\\
C^{ijkl}_0&=&-(i)^{\theta_i+\theta_j}(-1)^{\theta_{\tw_k}}
\sin\beta\cos\beta
\left (A^{(i)}_{k1}A^{(j)}_{k2}-A^{(i)}_{k2}A^{(j)}_{k1}\right ),\\
A^{(i)}_{11}&=&\frac{1}{\sqrt2}\left (gv^{(i)}_3+g^\prime v^{(i)}_4\right )
\cos\gamma_R-gv^{(i)}_1\sin\gamma_R,\\
A^{(i)}_{21}&=&\theta_y\left (-\frac{1}{\sqrt2}\left (gv^{(i)}_3+g^\prime 
v^{(i)}_4\right )\sin\gamma_R-gv^{(i)}_1\cos\gamma_R\right ),\\
A^{(i)}_{12}&=&\frac{1}{\sqrt2}\left (gv^{(i)}_3+g^\prime v^{(i)}_4\right )
\cos\gamma_L+gv^{(i)}_2\sin\gamma_L,\\
A^{(i)}_{22}&=&\theta_x\left (-\frac{1}{\sqrt2}\left (gv^{(i)}_3+g^\prime 
v^{(i)}_4\right )\sin\gamma_L+gv^{(i)}_2\cos\gamma_L\right ).\\
\end{eqnarray*}
Finally, 
\begin{equation}
g_{mnop}=-\frac{e}{(4\pi)^2}\left (m_{\tilde Z_i}C^{mnop}_i K
+m_{\tilde Z_j}C^{mnop}_j (I^2-K)+m_{\tw_k}C^{mnop}_0 I\right ).
\end{equation}
Here $C^{mnop}_I $ ($I=i,j,0$) are obtained from 
$C^{ijkl}_I$ by making 
the replacement $\sin\beta\rightarrow-\cos\beta, 
\cos\beta\rightarrow\sin\beta$.
In the above, $\gamma_L$ and $\gamma_R$ are chargino mixing angles
as defined by Tata; in that work\cite{reviews}, the $\theta_x$, $\theta_y$, 
$\theta_i$ and $\theta_{\tw_i}$ are also defined. 

\section*{Acknowledgments}
 
We thank Xerxes Tata for discussions.
This research was supported in part by the U.S. Department of Energy
under contract number DE-FG02-97ER41022.
	
% ---- Bibliography ----
%

\end{document}